\documentclass[aps,preprint,showpacs]{revtex4}%
\usepackage{amssymb}
\usepackage{amsfonts}
\usepackage{amsmath}
\usepackage{graphicx}%
\providecommand{\U}[1]{\protect\rule{.1in}{.1in}}
\begin{document}
\title[Hamiltonian's relativity]{Basic quantum Hamiltonian's relativistic corrections}
\author{Gintautas P. Kamuntavi\v{c}ius}
\affiliation{Physics Department, Vytautas Magnus University, Vileikos 8, Kaunas 44404, Lithuania}
\email{g.kamuntavicius@gmf.vdu.lt}

\begin{abstract}
After analyzing Dirac's equation, one can suggest that a well-known
quantum-mechanical momentum operator is associated with relativistic momentum,
rather than with non-relativistic one$.$ Consideration of relativistic energy
and momentum expressions allows us to define the non-relativistic,
relativistic and pseudo-relativistic (present in Schr\"{o}dinger equation)
kinetic energy operators. Consequences of kinetic energy operator's correction
for spectra of basic quantum Hamiltonians are investigated. In some cases this
correction can produce remarkable spectra modifications.

\end{abstract}
\date{2013.03.28}
\startpage{1}

\maketitle

\section{Introduction}

The many-particle Schr\"{o}dinger equation is the best tool for quantum
systems low-energy phenomena investigation. The existing powerful approximate
methods enable successful description of such systems as electrons in solid
state, molecules, atoms and even the atomic nuclei. The Hamiltonian is defined
as a non-relativistic operator, hence relativistic modifications are among the
most important. However, the solution of this problem is very problematic due
to the complex relativistic kinetic energy operator and the undefined
many-particle Dirac equation. This problem has been solved only in weakly
relativistic approximation for systems with Coulomb potential, such as an
atom. The main points of this solution are based on Hamiltonian's corrections,
saving the usual form of kinetic energy operator and introducing new
potentials like spin-orbit, following from corresponding Dirac equation. For
systems with strong interaction these Hamiltonian corrections are not
sufficient in order to obtain acceptable result. New ways need to be
investigated in order to solve this problem.

Let us take a new look at relativistic modifications for basic
quantum-mechanical Hamiltonians describing particle, moving in
spherically-symmetric external fields, created by deep well, harmonic
oscillator and Coulomb potentials. These Hamiltonians consist of one-particle
kinetic energy operator and external field potential operator, and can be
presented as:%
\begin{equation}
\mathbf{h}_{0}=-\frac{\left(  \hbar c\right)  ^{2}}{2mc^{2}}\Delta
+\mathbf{v}\left(  r,\tau\right)  , \label{OPHam}%
\end{equation}
where $mc^{2}$\ is particle's rest energy, given in $GeV,$ $\hbar c=0.197$
$GeV$ $fm$. $r\ $is absolute value of it's radius-vector and $\tau$ marks the
set of intrinsic variables, such as mass, charge, spin, isospin and possible
others.\ Such operators are the essential parts of many-particle Hamiltonian,
hence relativistic corrections are necessary for further investigation of this problem.

\section{Dirac equation transformations}

Dirac equation \cite{Dirac}, \cite{Davydov} is the basic for Hamiltonian's
relativistic corrections:
\begin{equation}
\left(  \mathbf{1}\frac{1}{c}\frac{\partial}{\partial t}+\sum_{k=1}%
^{3}\boldsymbol{\alpha}_{k}\frac{\partial}{\partial x_{k}}+\boldsymbol{\beta
}\frac{imc^{2}}{\hbar c}\right)  \Psi=0, \label{Dirac}%
\end{equation}
where $\boldsymbol{\alpha}_{k}$ and $\boldsymbol{\beta}$ are Dirac matrices
and wave-function $\Psi$ is defined as set of four functions - components.
Modified stationary equation for free particle with spin, equal $\hbar/2$ can
be written in Pauli matrices as system of two equations:
\begin{equation}
\left\{
\begin{array}
[c]{c}%
\left(  \boldsymbol{\varepsilon}-mc^{2}\right)  \varphi+i\hbar c\left(
\boldsymbol{\sigma}\cdot\boldsymbol{\nabla}\right)  \chi=0,\\
\left(  \boldsymbol{\varepsilon}+mc^{2}\right)  \chi+i\hbar c\left(
\boldsymbol{\sigma}\cdot\boldsymbol{\nabla}\right)  \varphi=0.
\end{array}
\right.  \label{Dirac S}%
\end{equation}
Here $\boldsymbol{\varepsilon}$ is relativistic particle's energy operator,
and $\boldsymbol{\sigma}$ - spin operator. $\varphi$ and $\chi$ are
two-component spinors composing entire wave-function
\begin{equation}
\Psi=\left(
\begin{array}
[c]{c}%
\varphi\\
\chi
\end{array}
\right)  .
\end{equation}
Let us define $\Psi$ as eigenfunction of operator $\boldsymbol{\varepsilon}$
corresponding to eigenvalue, equal $\varepsilon.$ Having in mind that
commutator of operators $\boldsymbol{\varepsilon}$ and $\boldsymbol{\nabla}$
equals zero, the eigenfunction of $\boldsymbol{\varepsilon}$ is also
eigenfunction of momentum operator $\mathbf{p=-}i\hbar\boldsymbol{\nabla}$
with eigenvalue, equal $p,$ one obtains that%
\begin{equation}
\varepsilon^{2}=\left(  mc^{2}\right)  ^{2}+\left(  pc\right)  ^{2}=\left(
\gamma mc^{2}\right)  ^{2}. \label{Epsilon}%
\end{equation}
This is the well-known expression for energy of the relativistic particle with
momentum $p=\gamma mv,$ where Lorentz factor%
\begin{equation}
\gamma=\left(  1-\left(  v/c\right)  ^{2}\right)  ^{-1/2}.
\end{equation}
This leads to the conclusion that quantum-mechanical momentum operator
$\mathbf{p=-}i\hbar\boldsymbol{\nabla}$ is associated with relativistic
momentum $p=\gamma mv$ rather than with non-relativistic one $p_{0}=mv.$

The non-relativistic momentum operator can be defined in a following way. From
Eq. $\left(  \ref{Epsilon}\right)  $, written in operators form, it follows
that
\begin{equation}
\mathbf{p}^{2}=\left(  \boldsymbol{\gamma}^{2}-1\right)  \left(  mc\right)
^{2},
\end{equation}
where the operator
\begin{equation}
\boldsymbol{\gamma}^{2}=\left(  1-\left(  \mathbf{p}_{0}/mc\right)
^{2}\right)  ^{-1}.
\end{equation}
Defining this classical Lorentz factor as quantum-mechanical operator, gives
us the possibility to introduce non-relativistic momentum operator,
non-relativistic, relativistic and pseudo-relativistic (present in
Schr\"{o}dinger equation) kinetic energy operators. As well as analyze
relations between them and investigate spectrum modifications when different
kinetic energy operators are present in Hamiltonian.

Applying the definition of the non-relativistic kinetic energy operator%
\begin{equation}
\mathbf{t}_{n}=\mathbf{p}_{0}^{2}/2m
\end{equation}
one obtains the expression:%
\begin{equation}
\mathbf{p}^{2}=\mathbf{p}_{0}^{2}\boldsymbol{\gamma}^{2}=2m\mathbf{t}%
_{n}\left(  1-2\mathbf{t}_{n}/mc^{2}\right)  ^{-1}.
\end{equation}
Let us introduce the operator present in Schr\"{o}dinger equation and call it
pseudo-relativistic kinetic energy operator:%
\begin{equation}
\mathbf{t}_{0}=\mathbf{p}^{2}/2m\equiv-\frac{\left(  \hbar c\right)  ^{2}%
}{2mc^{2}}\boldsymbol{\Delta}.
\end{equation}
It can be present in terms of $\mathbf{t}_{n}$ :
\begin{equation}
\mathbf{t}_{0}=\mathbf{t}_{n}\left(  1-2\mathbf{t}_{n}/mc^{2}\right)  ^{-1}.
\label{PperN}%
\end{equation}
From last expression it follows that
\begin{equation}
\mathbf{t}_{n}=\frac{\mathbf{p}^{2}}{2m}\left(  1+\left(  \mathbf{p/}%
mc\right)  ^{2}\right)  ^{-1}=\mathbf{t}_{0}\left(  1+2\mathbf{t}_{0}%
/mc^{2}\right)  ^{-1},
\end{equation}
or%
\begin{equation}
\mathbf{t}_{n}=\mathbf{t}_{0}\left[  1-2\mathbf{t}_{0}/mc^{2}+\left(
2\mathbf{t}_{0}/mc^{2}\right)  ^{2}-...\right]  . \label{TbarTnulis}%
\end{equation}
Therefore, the non-relativistic kinetic energy operator $\mathbf{t}_{n}$ is
far more complex than pseudo-relativistic one $\mathbf{t}_{0}.$

Having kinetic energy definition in classical relativity $t=\varepsilon
-mc^{2}$ in mind, one can present the system of Dirac equations $\left(
\ref{Dirac S}\right)  $ in the form of quantum-mechanical operators%
\begin{equation}
\left\{
\begin{array}
[c]{c}%
\mathbf{t}\varphi=c\left(  \boldsymbol{\sigma}\cdot\mathbf{p}\right)  \chi,\\
\left(  \mathbf{t}+2mc^{2}\right)  \chi=c\left(  \boldsymbol{\sigma}%
\cdot\mathbf{p}\right)  \varphi
\end{array}
\right.  \label{Dirac T}%
\end{equation}
and after usual manipulations%
\begin{equation}
\chi=\left(  \mathbf{t}+2mc^{2}\right)  ^{-1}c\left(  \boldsymbol{\sigma}%
\cdot\mathbf{p}\right)  \varphi,
\end{equation}%
\begin{equation}
\mathbf{t}\varphi=c\left(  \boldsymbol{\sigma}\cdot\mathbf{p}\right)  \left(
\mathbf{t}+2mc^{2}\right)  ^{-1}c\left(  \boldsymbol{\sigma}\cdot
\mathbf{p}\right)  \varphi,
\end{equation}
can obtain the relativistic kinetic energy operator:%
\begin{equation}
\mathbf{t}=\sqrt{\left(  mc^{2}\right)  ^{2}+\left(  \mathbf{p}c\right)  ^{2}%
}-mc^{2}.
\end{equation}
In terms of $\mathbf{t}_{0}$ it can be presented as%

\begin{equation}
\mathbf{t=}mc^{2}\left[  \sqrt{1+2\mathbf{t}_{0}/mc^{2}}-1\right]  .
\label{TT0}%
\end{equation}
This can be written as a series expansion:
\begin{equation}
\mathbf{t=t}_{0}\left[  1-\mathbf{t}_{0}/2mc^{2}+\left(  \mathbf{t}_{0}%
/mc^{2}\right)  ^{2}/2-...\right]  , \label{TTnulis}%
\end{equation}
or
\begin{equation}
\mathbf{t=t}_{0}\sum_{k=0}^{\infty}\frac{\left(  -1\right)  ^{k}}{2^{k}\left(
k+1\right)  }\left(
\begin{array}
[c]{c}%
2k\\
k
\end{array}
\right)  \left(  \mathbf{t}_{0}/mc^{2}\right)  ^{k}.
\end{equation}
The first term of all operators equals $\mathbf{t}_{0},$ see $\left(
\ref{TbarTnulis}\right)  $, $\left(  \ref{TTnulis}\right)  $, hence at
non-relativistic energies, when $t_{0}\ll mc^{2},$ all operators have the same form.

The expression of relativistic kinetic energy operator in terms of
non-relativistic one is%
\begin{equation}
\mathbf{t/}mc^{2}=\left(  1-2\mathbf{t}_{n}/mc^{2}\right)  ^{-1/2}-1.
\label{RperN}%
\end{equation}

The commutators of all pairs of presented kinetic energy operators equal zero,
all of them are differential operators of the same argument - radius vector of
particle - hence relations between eigenvalues can be defined as relations
between operators, and can be considered in terms of classic relativity. The
eigenvalues, corresponding to the same eigenfunction, are distributed in a
following way:%
\begin{equation}
t_{n}<t<t_{0}.
\end{equation}
Thus, the pseudo-relativistic operator produces the largest eigenvalue in
comparison with other two for the same eigenfunction. All operators are
positively defined, so it follows that at non-relativistic kinetic energy,
corresponding to the maximum possible velocity $v=c,$ when $t_{n}/mc^{2}=1/2,$
both (relativistic $\left(  \ref{RperN}\right)  $ and pseudo-relativistic
$\left(  \ref{PperN}\right)  $) kinetic energy operator's eigenvalues
approach, as necessary, the infinite values.

Therefore, the Schr\"{o}dinger equation is not completely non-relativistic.
Some relativistic corrections have already been taken into account applying
relativistic momentum operator instead of the non-relativistic one. In first
approximation, i.e. at negligible in comparison with $mc^{2}$ kinetic energy,
all operators coincide.

Dirac equation $\left(  \ref{Dirac S}\right)  $ can be transformed into
Schr\"{o}dinger equation form also for particle moving in external field,
which scalar potential is $\mathbf{v}.$ It is necessary to introduce the
expression $\mathbf{t=h-v,}$ where $\mathbf{h}$ is the Hamiltonian instead of
kinetic energy operator $\mathbf{t}$. The equation $\left(  \ref{Dirac T}%
\right)  $ can be rewritten in following form%
\begin{equation}
\mathbf{h}\left(
\begin{array}
[c]{c}%
\varphi\\
\chi
\end{array}
\right)  =\left(
\begin{array}
[c]{cc}%
\mathbf{v} & c\left(  \boldsymbol{\sigma}\cdot\mathbf{p}\right) \\
c\left(  \boldsymbol{\sigma}\cdot\mathbf{p}\right)  & \mathbf{v-}2mc^{2}%
\end{array}
\right)  \left(
\begin{array}
[c]{c}%
\varphi\\
\chi
\end{array}
\right)  .
\end{equation}
Now let's assume that present wave-function is an eigenfunction of Hamiltonian%
\begin{equation}
\mathbf{h}\left(
\begin{array}
[c]{c}%
\varphi\\
\chi
\end{array}
\right)  =e\left(
\begin{array}
[c]{c}%
\varphi\\
\chi
\end{array}
\right)  ,
\end{equation}
where $e$ is corresponding eigenvalue. The system of equations now is:%
\begin{equation}
\left\{
\begin{array}
[c]{c}%
\left(  e\mathbf{-v}\right)  \varphi=c\left(  \boldsymbol{\sigma}%
\cdot\mathbf{p}\right)  \chi,\\
\left(  e-\mathbf{v}+2mc^{2}\right)  \chi=c\left(  \boldsymbol{\sigma}%
\cdot\mathbf{p}\right)  \varphi.
\end{array}
\right.
\end{equation}
As usual, the following operation eliminates the wave-function component
$\chi.$ The equation for component $\varphi$ appears in form:%
\begin{equation}
e\varphi=\mathbf{v}\varphi+c\left(  \boldsymbol{\sigma}\cdot\mathbf{p}\right)
\left(  e-\mathbf{v}+2mc^{2}\right)  ^{-1}c\left(  \boldsymbol{\sigma}%
\cdot\mathbf{p}\right)  \varphi.
\end{equation}
The commutator of potential energy operator $\mathbf{v}$ and momentum operator
$\mathbf{p}$ does not equal zero. This makes the expression complex enough, so
it can be successfully developed only at fulfilled condition $e-\mathbf{v\ll}$
$2mc^{2},$ known as weakly relativistic approach \cite{QM}, \cite{Davydov}.
After the simple manipulations it can be presented as
\begin{equation}
\mathbf{h=h}_{0}\mathbf{+v}_{T}+\mathbf{v}_{S}+\mathbf{v}_{D},
\end{equation}
where apart the standard Hamiltonian $\mathbf{h}_{0}=\mathbf{t}_{0}%
\mathbf{+v,}$ the additional potentials appear.
\begin{equation}
\mathbf{v}_{T}=-\frac{\left(  e\mathbf{-v}\left(  r\right)  \right)  ^{2}%
}{2mc^{2}},
\end{equation}
equals relativistic kinetic energy second order term $\left(  \ref{TTnulis}%
\right)  ,$ converted to potential operator form.
\begin{equation}
\mathbf{v}_{S}=\frac{1}{2}\left(  \frac{\hbar c}{mc^{2}}\right)  ^{2}\left(
\mathbf{s}\cdot\mathbf{l}\right)  \frac{1}{r}\frac{d\mathbf{v}\left(
r\right)  }{dr}%
\end{equation}
is the spin-orbit potential (here $\mathbf{s}$ and $\mathbf{l}$ are
dimensionless spin and orbital momenta operators) and
\begin{equation}
\mathbf{v}_{D}=\frac{1}{8}\left(  \frac{\hbar c}{mc^{2}}\right)  ^{2}%
\Delta\mathbf{v}\left(  r\right)
\end{equation}
is the Darwin potential.

This investigation shows that potential's weakly relativistic modification has
to be considered only as some approximation for complete kinetic energy
conversion to relativistic form.

Dirac equation for particle, moving in external field, transformation into
Schr\"{o}dinger equation form was very attractive years ago. Namely because
kinetic energy expression is like one, present in Schr\"{o}dinger equation,
$e$ is characteristic for Schr\"{o}dinger equation energy of particle's bound
state in given potential (equal binding energy with negative sign) and
additional terms of a potential have a well-defined structure, appearing
because of the nonzero spin and associated magnetic momentum.

However nucleons or constituent quarks besides of spin have additional
intrinsic degrees of freedom, such as isospin and color and the realistic
potentials contain all operators, allowed by symmetry considerations, these
weakly relativistic modifications of strong interaction potentials are not necessary.

\section{Basic quantum Hamiltonian's relativity}

Let us now investigate bound-states spectrum of basic quantum-mechanical
Hamiltonian $\left(  \ref{OPHam}\right)  $ when relativistic kinetic energy
operator is applied instead of pseudo-relativistic one. The Schr\"{o}dinger
equation for these Hamiltonians is%
\begin{equation}
\left[  -\frac{\left(  \hbar c\right)  ^{2}}{2mc^{2}}\Delta+\mathbf{v}\left(
r,\tau\right)  -e\right]  \varphi\left(  \mathbf{r}\right)  =0.
\end{equation}
The spherically-symmetric potential provides the possibility to reduce this
equation to ordinary differential equation for radial function, defined in a
following way:%
\begin{equation}
\varphi\left(  \mathbf{r}\right)  =R_{nl}\left(  r\right)  Y_{l\mu}\left(
\vartheta\varphi\right)  ,
\end{equation}
where $Y_{l\mu}\left(  \vartheta\varphi\right)  $ is spherical harmonic and
$n=1,2,...$ equals the number of bound state, counted from the lowest one. The
radial equation after some transformations can be presented as%
\begin{equation}
\left[  \frac{1}{r}\frac{d^{2}}{dr^{2}}r-\frac{l\left(  l+1\right)  }{r^{2}%
}+\frac{2mc^{2}}{\left(  \hbar c\right)  ^{2}}\left(  e_{nl}-\mathbf{v}\left(
r\right)  \right)  \right]  R_{nl}\left(  r\right)  =0.
\end{equation}
Here the function $u_{nl}\left(  r\right)  =rR_{nl}\left(  r\right)  $ is more
acceptable than the radial function. Radial Schr\"{o}dinger equation for this
function has the form
\begin{equation}
\left[  \frac{d^{2}}{dr^{2}}-\frac{l\left(  l+1\right)  }{r^{2}}+\frac
{2mc^{2}}{\left(  \hbar c\right)  ^{2}}\left(  e_{nl}-\mathbf{v}\left(
r\right)  \right)  \right]  u_{nl}\left(  r\right)  =0.
\end{equation}
The attracting points of this radial function are as follows: it is normalized
by condition%
\begin{equation}
\int_{0}^{\infty}\left\vert u_{nl}\left(  r\right)  \right\vert ^{2}dr=1,
\end{equation}
so $\left\vert u_{nl}\left(  r\right)  \right\vert ^{2}$ equals the radial
probability density. Moreover, this function satisfies a condition%
\begin{equation}
\lim_{r\rightarrow0}u_{nl}\left(  r\right)  =0. \label{Bound0}%
\end{equation}
Next, the equation is transformed into dimensionless form. Introducing
dimension-free energy and potential, one obtains an equation, \ where each
member has dimension that equals $1/\left(  length\right)  ^{2}$ :%
\begin{equation}
\left[  \frac{d^{2}}{dr^{2}}-\frac{l\left(  l+1\right)  }{r^{2}}+2\frac
{mc^{2}\epsilon}{\left(  \hbar c\right)  ^{2}}\frac{\left(  e_{nl}%
-\mathbf{v}\left(  r\right)  \right)  }{\epsilon}\right]  u_{nl}\left(
r\right)  =0,
\end{equation}
where $\epsilon$ is combination of Hamiltonian parameters with energy
dimension. For deep well and Coulomb potentials this parameter equals
$mc^{2}.$ For Harmonic Oscillator potential
\begin{equation}
\mathbf{v}\left(  r\right)  =m\omega^{2}r^{2}/2
\end{equation}
parameter $\epsilon=\hbar\omega$ is the best choice. It follows that parameter
with length dimension for our problem is%
\begin{equation}
b^{2}=\frac{\left(  \hbar c\right)  ^{2}}{mc^{2}\epsilon}.
\end{equation}
Taking new radial variable
\begin{equation}
\xi=r/b,
\end{equation}
the equation converts to form%
\begin{equation}
\left[  \frac{d^{2}}{d\xi^{2}}-\frac{l\left(  l+1\right)  }{\xi^{2}}%
+2\frac{\left(  e_{nl}-\mathbf{v}\left(  b\xi\right)  \right)  }{\epsilon
}\right]  u_{nl}\left(  \xi\right)  =0. \label{SEbedim}%
\end{equation}
The discrete spectrum eigenfunctions of this equation form the orthonormal
set:
\begin{equation}
\int_{0}^{\infty}u_{n^{\prime}l}^{+}\left(  \xi\right)  u_{nl}\left(
\xi\right)  d\xi=\delta_{n^{\prime},n}. \label{Ortonorm}%
\end{equation}
The kinetic energy expectation value is%
\begin{equation}
\left\langle \mathbf{t}_{0}\right\rangle _{nl}=-\frac{1}{2}\epsilon\int%
_{0}^{\infty}u_{nl}^{+}\left(  \xi\right)  \left[  \frac{d^{2}}{d\xi^{2}%
}-\frac{l\left(  l+1\right)  }{\xi^{2}}\right]  u_{nl}\left(  \xi\right)
d\xi.
\end{equation}
The potential energy expectation value equals%
\begin{equation}
\left\langle \mathbf{v}\right\rangle _{nl}=\int_{0}^{\infty}u_{nl}^{+}\left(
\xi\right)  \mathbf{v}\left(  b\xi\right)  u_{nl}\left(  \xi\right)  d\xi.
\end{equation}

Let us now compare bound eigenvalues spectra of two Hamiltonians - one with
pseudo-relativistic kinetic energy operator, i.e.%
\begin{equation}
\mathbf{h}_{0}=\mathbf{t}_{0}+\mathbf{v}%
\end{equation}
and second - with relativistic kinetic energy operator%
\begin{equation}
\mathbf{h}=\mathbf{t}+\mathbf{v,}%
\end{equation}
defined in $\left(  \ref{TT0}\right)  $. Let us mark the eigenvalue of the
first Hamiltonian as $e_{nl}.$ The eigenvalue of the second Hamiltonian in
first perturbation theory approximation equals%
\begin{equation}
\left\langle \mathbf{h}\right\rangle _{nl}=e_{nl}+\left\langle \Delta
\mathbf{h}\right\rangle _{nl},
\end{equation}
where, according to given above expressions, the eigenvalue correction due to
kinetic energy relativistic expression is
\begin{equation}
\left\langle \Delta\mathbf{h}\right\rangle _{nl}=\left\langle \mathbf{t-t}%
_{0}\right\rangle _{nl}=mc^{2}\left[  \sqrt{1+2\left\langle \mathbf{t}%
_{0}\right\rangle _{nl}/mc^{2}}-1-\left\langle \mathbf{t}_{0}\right\rangle
_{nl}/mc^{2}\right]  \label{Korekcija}%
\end{equation}

The first basic potential, which is going to be analyzed, is spherically
symmetric well of infinite depth. This potential describes free particle,
moving inside of sphere with no probability to find particle penetrated to
outer region and equals:%
\begin{equation}
\mathbf{v}\left(  r\right)  \mathbf{=}\left\{
\begin{array}
[c]{c}%
0,\quad if\quad r\leq d\\
\infty,\quad if\quad r>d
\end{array}
\right.  ,
\end{equation}
where $d$ is sphere radius. The solution of corresponding Schr\"{o}dinger
equation $\left(  \ref{SEbedim}\right)  $ at $\mathbf{v}\left(  b\xi\right)
\equiv0$ is%
\begin{equation}
u_{nl}\left(  \xi\right)  =\xi j_{l}\left(  \xi\right)  ,
\end{equation}
where $j_{l}\left(  \xi\right)  $ is the spherical Bessel function (\cite{AS},
\S \ 10.1). The quantum number $n=1,2,...$ is defined according to the
boundary condition
\begin{equation}
u_{nl}\left(  d/b\right)  =0.
\end{equation}
This equation has an infinite number of solutions, defined by different values
of $b=b_{n}.$ For $l=0$
\begin{equation}
u_{n0}\left(  \xi\right)  =N_{n0}\sin\xi,
\end{equation}
hence%
\begin{equation}
b_{n}=d/n\pi,\qquad N_{n0}=\sqrt{2/n\pi}.
\end{equation}
Thus, the eigenvalues of Schr\"{o}dinger equation $\left(  \ref{SEbedim}%
\right)  $ equal to:%
\begin{equation}
e_{n0}/mc^{2}=z\left(  n\pi\right)  ^{2}/2,\label{Eigen1}%
\end{equation}
where $z$ is the dimensionless well parameter%
\begin{equation}
z=\left(  \frac{\hbar c}{d\text{ }mc^{2}}\right)  ^{2}.
\end{equation}
Potential equals zero in an inside region, hence the kinetic energy
expectation value for this potential equals eigenvalue $e_{n0}$. The
perturbation due to relativistic corrections is%
\begin{equation}
\left\langle \Delta\mathbf{h}\right\rangle _{n0}/mc^{2}=\sqrt{1+z\left(
n\pi\right)  ^{2}}-1-z\left(  n\pi\right)  ^{2}/2.
\end{equation}
This equals one percent of eigenvalue $\left(  \ref{Eigen1}\right)  $ at
\begin{equation}
z\left(  n\pi\right)  ^{2}\approx0.083.
\end{equation}
For nucleon $\left(  mc^{2}\approx1\text{ }GeV\right)  $, moving in well,
which radius equals $1$ $fm,$ the parameter
\begin{equation}
z\approx\left(  \frac{0.2\text{ }GeV\text{ }fm}{1\text{ }fm\ 1Gev}\right)
^{2}=0.04,
\end{equation}
so this kinetic energy correction give a remarkable effect even for ground
state $\left(  n=1\right)  $ of corresponding strong process model. After
kinetic energy correction, the spectrum of spherical well will change in
comparison with the well-known levels distribution, whose energies are
proportional to $n^{2}.$ The distances between pairs of neighboring levels
will be smaller than in spectrum without relativistic kinetic energy correction.

Three-dimensional Harmonic Oscillator potential is next on the list. As
previously defined, the parameters of the equation equal:%
\begin{equation}
\epsilon=\hbar\omega,\qquad\epsilon b^{2}=\left(  \hbar c\right)  ^{2}%
/mc^{2},\qquad b^{2}=\hbar/m\omega.
\end{equation}
So,%
\begin{equation}
\mathbf{v}\left(  b\xi\right)  =\epsilon\xi^{2}/2,
\end{equation}
and the dimensions-free equation is%
\begin{equation}
\left[  \frac{d^{2}}{d\xi^{2}}-\frac{l\left(  l+1\right)  }{\xi^{2}}-\xi
^{2}+\frac{2e_{nl}}{\hbar\omega}\right]  u_{nl}\left(  \xi\right)  =0.
\end{equation}
The eigenvalues of this equation are%
\begin{equation}
e_{nl}=\hbar\omega\left(  2n+l-1/2\right)  .
\end{equation}
The eigenfunctions, normalized by condition $\left(  \ref{Ortonorm}\right)  $
are defined as in (\cite{AS}, \S \ 22.6.18):
\begin{equation}
u_{nl}\left(  \xi\right)  =N_{nl}\xi^{l+1}e^{-\xi^{2}/2}L_{n-1}^{l+1/2}\left(
\xi^{2}\right)  .
\end{equation}
Here normalization constant%
\begin{equation}
N_{nl}^{2}=\frac{2^{n+l+1}\left(  n-1\right)  !}{\sqrt{\pi}\left[  2\left(
n+l\right)  -1\right]  !!},
\end{equation}
Laguerre polynomial is defined as%
\begin{equation}
L_{k}^{\lambda}\left(  x\right)  =\sum_{j=0}^{k}\left(  -1\right)  ^{j}\left(
\begin{array}
[c]{c}%
k+\lambda\\
k-j
\end{array}
\right)  \frac{x^{j}}{j!},
\end{equation}
where%
\begin{equation}
\left(
\begin{array}
[c]{c}%
\beta\\
j
\end{array}
\right)  =\frac{\Gamma\left(  \beta+1\right)  }{j!\Gamma\left(  \beta
-j+1\right)  }%
\end{equation}
is a binomial coefficient and $\Gamma\left(  \beta\right)  -$ Gamma function
$\left[  \Gamma\left(  \beta+1\right)  =\beta\Gamma\left(  \beta\right)
,\Gamma\left(  1/2\right)  =\sqrt{\pi}\right]  $. The potential expectation
value equals the kinetic energy expectation value and both are defined as%
\begin{equation}
\left\langle \mathbf{v}\right\rangle _{nl}/mc^{2}=\left\langle \mathbf{t}%
_{0}\right\rangle _{nl}/mc^{2}=e_{nl}/2mc^{2}=z\left(  2n+l-1/2\right)  /2,
\end{equation}
where dimensionless parameter
\begin{equation}
z=\left(  \frac{\hbar c}{b\text{ }mc^{2}}\right)  ^{2}=\hbar\omega/mc^{2}.
\end{equation}
The Hamiltonian eigenvalue relativistic correction, defined in $\left(
\ref{Korekcija}\right)  $ is:%
\begin{equation}
\left\langle \Delta\mathbf{h}\right\rangle _{nl}/mc^{2}=\sqrt{1+z\left(
2n+l-1/2\right)  }-1-z\left(  2n+l-1/2\right)  /2.
\end{equation}
As in square well case, this correction again will produce one percent of
bound state energy at%
\begin{equation}
z\left(  2n+l-1/2\right)  \approx0.083.
\end{equation}
As in earlier evaluation, at $z\approx0.04,$ this correction will be
remarkable even in ground state of strong interaction models with such
Harmonic Oscillator potential, i.e. at $n=1,$ $l=0.$ The spectrum of corrected
Hamiltonian will not consist of equidistant levels. The levels will be
distributed closer than in the well-known Harmonic Oscillator spectrum with
usual pseudo-relativistic kinetic energy operator.

The last basic Hamiltonian is for Hydrogen-like atom with infinitely heavy
nucleus. The Coulomb potential equals%
\begin{equation}
\mathbf{v}\left(  r\right)  =-\alpha Z\hbar c\frac{1}{r},
\end{equation}
where $\alpha\approx1/137.036$ is fine structure constant, $Ze$ - nucleus
charge and $\hbar c\approx0.197$ $keV$ $nm.$ The length parameter of this
problem equals Bohr radius:%
\begin{equation}
a=\frac{\hbar c}{\alpha Zmc^{2}},
\end{equation}
the dimensionless problem parameter%
\begin{equation}
z=\left(  \frac{\hbar c}{a\cdot mc^{2}}\right)  ^{2}=\left(  \alpha Z\right)
^{2}.
\end{equation}
After the new variable definition%
\begin{equation}
\xi=r/a,
\end{equation}
the dimensions-free equation looks as%
\begin{equation}
\left[  \frac{d^{2}}{d\xi^{2}}-\frac{l\left(  l+1\right)  }{\xi^{2}}+\frac
{2}{\xi}+\frac{2e_{nl}}{zmc^{2}}\right]  u_{nl}\left(  \xi\right)  =0.
\end{equation}
The eigenvalue equals%
\begin{equation}
e_{nl}/mc^{2}=-z/2n^{2}.
\end{equation}
The corresponding eigenfunction is (\cite{AS}, \S \ 22.6.17):%
\begin{equation}
u_{nl}\left(  \xi\right)  =N_{nl}\rho^{l+1}e^{-\rho/2}L_{n-l-1}^{2l+1}\left(
\rho\right)  ,
\end{equation}
where%
\begin{equation}
\rho=2\xi/n
\end{equation}
and normalization factor%
\begin{equation}
N_{nl}^{2}=\frac{\left(  n-l-1\right)  !}{n^{2}\left(  n+l\right)  !}.
\end{equation}
The potential energy expectation value for this problem is two times larger
than the eigenvalue%
\begin{equation}
\left\langle \mathbf{v}\right\rangle _{nl}/mc^{2}=2e_{nl}/mc^{2}=-z/n^{2},
\end{equation}
so the kinetic energy mean value is%
\begin{equation}
\left\langle \mathbf{t}_{0}\right\rangle _{nl}/mc^{2}=-e_{nl}/mc^{2}=z/2n^{2}.
\end{equation}

The expression for eigenvalue correction now is:%
\begin{equation}
\left\langle \Delta\mathbf{h}\right\rangle _{nl}/mc^{2}=\sqrt{1+z/n^{2}%
}-1-z/2n^{2}.
\end{equation}

This equals one percent of energy eigenvalue when%
\begin{equation}
Z/n\approx39.48.
\end{equation}
It happens only at extremely hard conditions. It reaches maximum value for
ground state only at large nucleus charge. This corresponds to the known
character of relativistic effects, significant mainly for heavy atoms
(\cite{Levine}, \S \ 11.7).

\section{Conclusions}

Our consideration of relativistic corrections for basic quantum Hamiltonian
has shown that the well-known momentum operator is not completely
non-relativistic, as it has been widely believed. Pseudo-relativistic kinetic
energy operator, present in Schr\"{o}dinger equation produces larger kinetic
energy expectation values than the relativistic operator. It appears that
relativistic kinetic energy correction can produce remarkable Schr\"{o}dinger
equation spectrum transformations. The further consideration of Dirac
equations transformation into Schr\"{o}dinger form has shown that kinetic
energy modification, known as weakly relativistic potential correction, can
not be useful for actual problems, because modern strong interaction realistic
potentials have already all the possible potential members, allowed by
symmetry consideration.

\end{document}